\documentclass[usegraphicx]{mn2e}
\begin{document}
\title[Incongruity of the unified scheme]
{Incongruity of the unified scheme with a 3CRR-like equatorial strong-source sample}
\author[Singal \&  Singh]{Ashok K. Singal \& Raj Laxmi Singh\\
{Astronomy and Astrophysics Division, Physical Research Laboratory, 
Navrangpura, Ahmedabad - 380 009, India}\\
\\E-mail: asingal@prl.res.in}

\date{Accepted . Received ; in original form }
\maketitle

\begin{abstract}
We examine the consistency of the unified scheme of the powerful extragalactic radio sources with the  
408 MHz BRL sample from the equatorial sky region, selected at the same flux-density level as the 3CRR sample. 
We find that, unlike in the 3CRR sample, a  
foreshortening in the observed sizes of quasars, expected from the orientation-based unified scheme model, is not seen 
in the BRL sample, at least in different redshift bins up to $z \sim 1$. Even the quasar fraction in individual 
redshift bins up to $z \sim 1$ does not match with that expected from the unified scheme, where radio galaxies and quasars 
are supposed to belong to a common parent population at all redshifts. This not only casts strong doubts on the unified scheme, 
but also throws up an intriguing result that in a  
sample selected from the equatorial sky region, using almost the same criteria as in the 3CRR sample from the 
northern hemisphere, the relative distribution of radio galaxies and quasars differs qualitatively 
from the 3CRR sample.
\end{abstract}
\section{Introduction}
In the currently popular orientation-based unified scheme (OUS),  
due to geometrical projection, the observed sizes of quasars 
should appear smaller than those of radio galaxies (RGs).  
Such a thing was noticed by Barthel (1989) in the redshift range 
$0.5<z<1$ in the 3CRR (3rd Cambridge twice revised) sample (Laing et al. 1983), based on which he proposed OUS.  
He suggested that both RGs and quasars belong to the same
parent population of radio sources, and that a source appears as
a quasar only when its major radio-axis happens to be oriented within a certain critical   
angle ($\xi_{c}$) around the observer's line of sight. 
Taking a cue from the case of the radio galaxy 3C234 and the Seyfert galaxy NGC1068 
(Antonucci 1984; Antonucci \& Miller 1985), Barthel proposed that   
the nuclear continuum and broad-line optical emission region 
is surrounded by an optically-thick torus and $\xi_{c}$ is the half cone-opening angle of the torus. 
In this model, in the case of RGs the observer's line of sight passes through the obscured region 
which hides the bright optical nucleus and the broad-line region. 
RGs and quasars are otherwise considered to be intrinsically indistinguishable 
and all differences in their observed radio properties  are 
attributed to their supposedly different orientations with respect
to the observer's line of sight; in particular, the observed smaller values
of radio sizes of quasars in Barthel's 3CRR sample were attributed to their larger
geometric projection effects because of the smaller inclinations
of their radio axes with respect to the observer's line of sight.
Recently in an about five times deeper MRC (Molonglo Reference Catalogue) sample (Kapahi et al. 1998a,b) such a difference 
in radio sizes of quasar and RGs was not seen (Singal \& Singh 2013), casting serious doubts on the 
unified scheme models.
Since the two samples (3CRR and MRC) have somewhat different flux-density limits, a question could arise that do 
two samples differing merely by a factor of five in flux-density, have radio source populations 
that differ so much that there could be such a  
large qualitative difference in the relative size distributions of RGs and quasars 
between the two samples? Or perhaps what is more pertinent here, what would have been the 
verdict if these two samples were at roughly a similar flux-density level?  We examine this issue here by using a sample,  
selected from the equatorial sky region, whose selection criteria are very similar to those of the 3CRR, and which 
presently is perhaps the best available sample to match the 3CRR sample. 

\section{The Source Sample}
For our investigations we have chosen the radio complete BRL sample (Best et al. 1999) at 408 MHz,  
selected from the Molonglo Reference Catalogue (Large et al. 1981) according 
to the criteria $S_{408} \geq 5$ Jy, $-30^{\circ}\leq\delta\leq10^{\circ}$  and $|b|\geq10^{\circ}$, and after rejecting the 
known galactic sources. The chosen flux-density limit of the sample when translated to 178 MHz using a typical spectral index of 0.85, 
corresponds roughly to the flux-density limit of the 3CRR sample. The sample is thus selected to have properties similar to 
the northern 3CRR sample, and to be visible to a combination of existing northern telescopes such as 
the VLA and large southern hemisphere telescope facilities. The sample is now 
100 per cent optically identified and redshifts are now available for all 178 sources (Best et al. 2000, 2003) in the sample     
that comprises 127 RGs, 49 quasars, one starburst galaxy and one Seyfert galaxy. From the overall redshift distributions, 
the BRL sample does look similar to the 3CRR sample (cf. Fig.  53 of Best et al. 1999). We have taken the radio and optical 
parameters for various sources from Table 2 of Best et al. (1999), where it has all the required information for the sample in a 
tabular form.
\begin{figure}
\scalebox{0.34}{\includegraphics{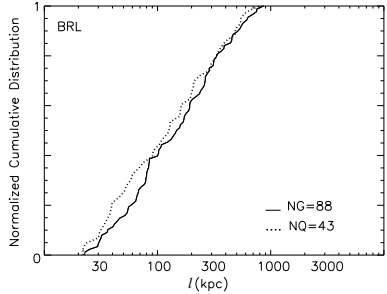}}
\caption{Normalized cumulative distributions of linear size ({\em l}) of RGs (continuous curves) and 
quasars (dotted curves) for the BRL sample. NG and NQ give the numbers of RGs and quasars respectively.}
\end{figure}

Among the RGs there are 14 broad-line radio galaxies (BLRGs). 
It is well known that the emission-line spectra for BLRGs
is very much similar to that of the distant quasars (see e.g. Osterbrock \& Mathews 1986)
and it may be more appropriate to treat BLRGs as quasars. 
Therefore we have clubbed BLRGs with quasars, making it 63 quasars and 113 RGs in the sample. 
The sample includes 14  sources (11 quasars and 3 RGs) with spectral index $\alpha \le 0.5$ (with $S\propto \nu^{-\alpha}$); 
we have excluded these flat-spectrum cases, 
since these mostly are core-dominant cases where the relativistic beaming might introduce serious selection effects.  
As only the powerful RGs are supposed to partake in unification with quasars, 
we have confined ourselves to only the strong sources with $P_{408} \ge 5 \times 10^{25}$ W Hz$^{-1}$ 
(for $H_{0}=71\,$km~s$^{-1}$\,Mpc$^{-1}$, $\Omega_m=0.27$, and 
$\Omega_{\Lambda}=0.73$; Spergel et al. 2003).  
This limit corresponds to the FR I/II luminosity break  
$P_{178} = 2 \times 10^{25}$ W Hz$^{-1}$ sr$^{-1}$ (for $H_{0}=50\,$km~s$^{-1}$\,Mpc$^{-1}$) 
of Fanaroff \& Riley (1974). There are 10 RGs below the FRI/II luminosity break, which we have excluded;   
the quasars in any case are all falling above this luminosity limit. 
Also in the BRL sample there is a large fraction ($\sim 0.15$) of compact steep spectrum sources 
(CSSS; linear size $< 20$ kpc), comprising 14 RGs and 9 quasars. The CSSS have their dominant radio emission on sub-galactic 
scales and thus might be a different class  
than the FRII class of sources whose unification is sought in OUS (Barthel 1989), and have therefore been excluded. 
Our final BRL sample then contains 131 sources, with 88 RGs and 43 quasars. For the 3CRR sample we have taken the 
optical and radio data from {\em https://www.astrosci.ca/users/willottc/3crr/3crr.html}, which maintains latest updates of the 
Laing et al. (1983) data. 
It should be noted that the steep spectrum quasars and FRII RGs almost always have 
edge-brightened radio morphologies with bright hot-spots near extremities, which makes it possible to define size of the radio 
source between the hot-spots, independent of the sensitivity of the observing telescope, both for the BRL and 3CRR samples. 
Since we intend to compare the size distributions in the BRL sample with those in the 3CRR sample, for uniformity sake 
we have applied exactly the same selection criteria for the 3CRR sample as well, whereby in the latter 
we get 130 sources, with 85 RGs and 45 quasars. 
\begin{figure}
\scalebox{0.34}{\includegraphics{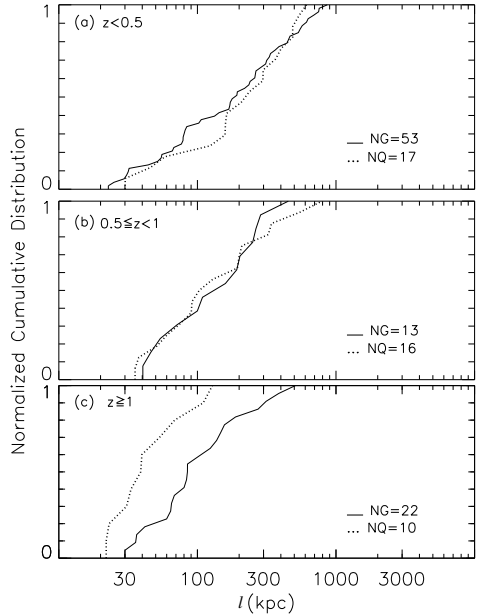}}
\caption{Normalized cumulative distribution of linear size ({\em l}) of RGs (continuous curves) and quasars (dotted curves)
excluding CSS sources, 
for the BRL sample in (a) $z < 0.5$, (b) $0.5\leq z<1$ and (c) $z \geq 1$ bins. 
NG and NQ give the numbers of RGs and quasars in each redshift bin.}
\end{figure}

\section{Results}
Fig.  1 shows a normalized cumulative distribution of the linear sizes of all RGs and quasars in the BRL sample. 
In the total BRL sample the fraction of quasars is $43/(88+43) \sim 0.33$, implying only about one third of the sources 
are quasars, similar to what was seen in the 3CRR sample of Barthel (1989), based on which OUS (with a cone opening angle 
$\xi_{c}\sim 45^{\circ}$), was proposed. In a picture consistent with OUS, BRL quasar sizes should also be statistically smaller 
than of RGs by a factor of two, but we  do not find it to be so (Fig.  1), in disagreement with OUS. 
Fig. 2 shows a normalized cumulative distribution of the linear sizes of RGs and quasars in three different redshift bins. 
The first thing we note from the figure is that the relative numbers of RGs and quasars as well as their radio sizes 
vary from one redshift bin to another. At low redshifts ($z<0.5$) number of quasars is relatively very small while 
in the intermediate redshift bin it is rather large.  The relative 
size distributions also show similar disparities. In the $z<0.5$ (Fig. 2a) and $0.5 \le z <1$ bins (Fig. 2b) quasars sizes do not 
appear in any 
way systematically smaller than those of RGs, thus negating the geometric-foreshortening hypothesis of OUS, however in the $z>1$ 
bin (Fig. 2c), the quasar sizes are a factor of two smaller than those of RGs. 

As there are a large number of CSSS cases ($ \sim 15\%$) in the BRL sample, in Fig. 3 we have included CSSS to check if some of 
unexpected results could be due to their presence. A comparison of Figs. 2 and 3 shows that there is almost no change in our above 
results, except perhaps in the intermediate redshift bin ($0.5 \le z <1$), where quasars appear slightly larger in sizes than RGs. 
In any case there is no evidence at least for $z<1$, that quasars sizes are smaller due to geometric projection according to OUS.  
\begin{figure}
\scalebox{0.34}{\includegraphics{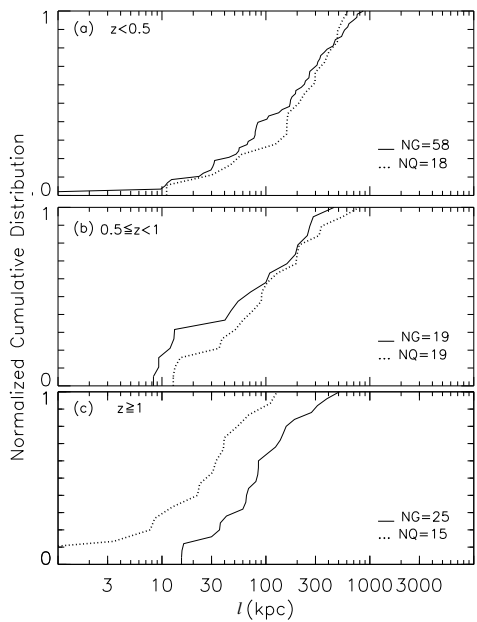}}
\caption{Normalized cumulative distribution of linear size ({\em l}) of RGs (continuous curves) and quasars (dotted curves), 
including CSS sources, 
for the BRL sample in (a) $z < 0.5$, (b) $0.5\leq z<1$ and (c) $z \geq 1$ bins. 
NG and NQ give the numbers of RGs and quasars in each redshift bin.}
\end{figure}

The results are summarized in Table 1, which is organized in the
following manner: 
(1) Sample used (BRL or 3CRR).
(2) Redshift bin.
(3) Number of RGs in that bin. 
(4) Number of quasars in that bin. 
(5) Fraction of quasars in that bin. 
(6) Median value (kpc) of size distribution for RGs. 
(7) Median value (kpc) of size distribution for quasars. 

From Table 1, we see that the relative number of RGs and quasars (or equivalently the quasar fraction $f_q$) 
in various redshift bins fluctuates significantly between the two samples. Quasar sizes are marginally larger than those of RGs in the 
lowest redshift bin ($z<0.5$) in both samples, but in the intermediate redshift bin, while 3CRR data show compatibility with OUS 
(not surprising since Barthel used only these data to propose OUS), in the BRL sample things appear different. Number of quasars 
seems more than of RGs and the size ratio is less than 1.4, while for the 3CRR case the size ratio in this bin is 2.4. 
In fact a Kolmogorov-Smirnov (K-S) test shows that it is only at a very low ($<10\%$) confidence level one could state that 
the two size distributions in the BRL sample in this redshift bin are different, in other words the two size 
distributions (Fig. 2b) are almost indistinguishable statistically. In the high redshift ($z>1$) bin the 
quasar sizes are about a factor of two or more smaller than those of RGs in both samples. It might sound in accordance with OUS, 
but the numbers of RGs and quasars are consistent with OUS only in the BRL sample. In the 3CRR case there are too many quasars 
(or too few RGs). But what is even more surprising is that the sizes of RGs and quasars in the BRL sample are a factor of  $\sim 2.5$  
smaller than those of 3CRR RGs and quasars in this high redshift bin.
\begin{table}
\caption{Numbers and median size values of radio galaxies and quasars in different redshift bins for the two samples.}
\begin{tabular}{@{}cccccccc}
\hline
Sample & Redshift bin & NG & NQ & $f_q$ & $l_{m}$(G) & $l_{m}$(Q)\\
(1)&\,(2)&(3)&(4)&(5)&(6)&(7)\\
\hline
BRL & $z<0.5$ & 53 & 17 & 0.24 & 192 & 238\\
3CRR & $z<0.5$ & 43 & 13 & 0.23 & 290 & 383\\\\
BRL & $0.5\leq z<1$ & 13 & 16 &  0.55 & 159 & 116\\
3CRR & $0.5\leq z<1$ & 27 & 14 & 0.34 & 260 & 110 \\\\
BRL & $z>1$ & 22 & 10 & 0.31 & 85 & 39 \\
3CRR & $z>1$ & 15 & 18 & 0.55 & 206 & 107\\
\hline
\end{tabular}
\end{table}
\section{Discussion}
It was only a limited redshift range ($0.5<z<1$) in the 3CRR sample that Barthel (1989) had investigated to propose  
OUS. Singal (1993) has subsequently shown that the data in other redshift bins in the 3CRR sample itself does not seem to 
conform to the OUS scenario. The situation looks particularly grave in the redshift range $z<0.5$, where 
according to Antonucci (2012) ``all hell breaks loose''. Not only is there 
a huge excess of RGs over the quasars, but also the quasar sizes are not very different from those of RGs.  
This has been later explained in the literature by proposing that  
there is a large population of low-excitation galaxies (LEGs),
which make a significant contribution to the number of FR~II-type radio galaxies 
at low redshifts (see e.g. Hine \& Longair 1979).
Laing et al. (1994) have pointed out that these optically dull LEGs are unlikely to appear
as quasars when seen end-on and that these should be excluded from the sample while testing the unified 
scheme models. From infrared observations also there is evidence of a population of powerful radio galaxies, concentrated at 
low redshifts, which lack the hidden quasar (see Antonucci 2012 and the references therein). 
Using both X-ray and Mid-IR data, Hardcastle et al. (2009) showed convincingly that almost all objects classed as LEGs
in optical spectroscopic studies lack a radiatively efficient active nucleus.   
Thus such a population of FR II RGs with no hidden quasars, 
concentrated only at low redshifts ($z\stackrel{<}{_{\sim}}0.5$)  
could make the apparently anomalous number and size distributions of RGs and quasars at low redshifts in the 3CRR sample 
(Singal 1993) consistent with OUS, and perhaps it might also hold true for the BRL sample. 
The overall redshift distribution was shown by Best et al (1999) to be  
similar to that of the 3CRR, with a large excess of RGs at low redshifts ($z<0.5$), which  
is consistent with a large population of LEGs at these redshifts. 
\begin{figure}
\scalebox{0.34}{\includegraphics{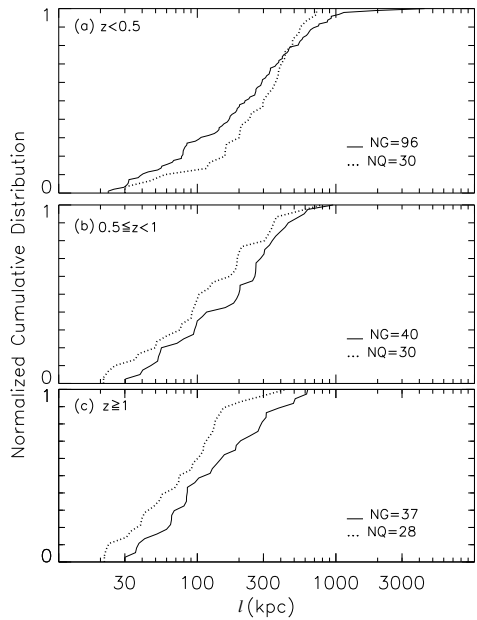}}
\caption{Normalized cumulative distribution of linear size ({\em l}) of RGs and quasars 
in three redshift bins for the combined (3CRR+BRL) sample. 
NG and NQ give the numbers of RGs and quasars in respective bin as indicated.}
\end{figure}

Therefore it is perhaps not that surprising that the quasar sizes do not seem smaller than those of the RGs in the $z<0.5$ bin 
as even the 3CRR data had shown this anomaly (Singal 1993). However, what is of more concern is that in the intermediate redshift 
bin ($0.5 \le z <1$) also the quasar numbers and sizes do not seem systematically smaller than of the RGs in the BRL sample 
(Fig. 2b). After all it was this redshift bin which Barthel (1989) had examined 
for the 3CRR sample and found quasar numbers as well as radio sizes to be about half of RGs and 
and that was the prime basis for the OUS hypothesis. However, in a matched independent BRL sample,  
the quasar numbers or sizes are not in the same way smaller than of RGs in this particular redshift bin. 
This is rather disconcerting, as the BRL sample is supposed to be very similar to the 3CRR one, with all selection 
criteria matched with those of the 3CRR, except that it is from a different (equatorial) sky region. 

We may add here that Best et al. (1999) have used a combined sample of BRL and 3CRR sources and reached a 
conclusion that the combined data are consistent with OUS. In fact they found quasars sizes to be about a factor of 
two smaller than those of RGs for $z>0.5$. To verify that we are not getting different result from 
them because of our somewhat different methodology, we used our procedure on the combined BRL+3CRR sample and the 
results are shown in Fig. 4. Of course in the low redshift bin ($z<0.5$) we find the same problem that 
quasars are larger in size than RGs. Best et al. (1999) had almost no quasars in this redshift bin (their Fig. 58) 
because they had included BLRGs with RGs 
while we have counted them with quasars (a better thing to do). But in other redshift bins our results for the combined sample 
are more or less consistent 
with Best et al. (apart from some differences due to BLRGs). It should be noted that in the intermediate redshift range 
($0.5 \le z <1$) much larger numbers of RGs in the 3CRR sample makes it dominant and therefore its influence on size 
distribution more pronounced in the combined sample. Also the quasar fraction in our combined sample ($\sim 0.43$) at $z>0.5$ 
is similar to that seen in Fig. 57 of Best et al. However when examined by itself alone, the BRL sample  
does seem to violate OUS in the $0.5 \le z<1$ bin.
\begin{figure}
\scalebox{0.34}{\includegraphics{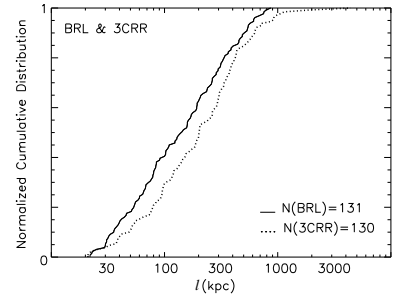}}
\caption{Normalized cumulative distribution of linear size ({\em l}) of radio sources 
for the 3CRR and BRL samples.}
\end{figure}
   
One could perhaps get away with the apparent anomaly in size distribution for both BRL and 3CRR samples in the redshift 
range $z < 0.5$ because it could be explained by 
a large number of LEGs there with relatively smaller radio sizes to offset the large sizes of 
high excitation RGs, and for which there is some evidence in the 3CRR case (Hardcastle et al. 1998). 
But how to do a similar thing at higher redshifts in the BRL sample? 
For example in the 3CRR data, while there are 16 LEGs out of total 43 radio galaxies in the $z < 0.5$ redshift bin, 
there is only one LEG found among 27 radio galaxies in the  $0.5<z<1$ redshift bin, which is insignificant to 
affect results either way. If we propose the presence of a large LEG population in the BRL sample that might exist at $0.5<z<1$ 
as well, we will not be still able to explain the observed anomaly. 
Such a proposal will go against OUS because already there is a lack of sufficient number of RGs in this redshift bin, and if 
some out of them are LEGs with no hidden quasars and thus not partaking in OUS, then the problem becomes even more acute.
 
To test that the BRL and 3CRR samples have statistically significant differences, we have performed the K-S  
test on their cumulative radio size distributions. 
Fig. 5 shows a plot of the cumulative radio size distributions for radio sources (RGs+quasars) in BRL and 3CRR samples.
The K-S test shows at a 92\% confidence level for them to be drawn from different populations.  
Perhaps it might be better to apply the test to RGs and quasars separately in the two samples. 
Fig. 6 shows cumulative radio size distributions separately for RGs and quasars 
in the two samples. When we apply the K-S test to cumulative radio size distributions of quasars alone, we get again 
similar result as above. However when we apply the K-S test to 
RGs alone, then we find the two distributions to be drawn from different populations at a confidence level of 99.7\%. 
This of course is very surprising as the two samples have been selected using almost the same criteria. 
In fact one wonders, for example, if one had the BRL sample available first 
(instead of the 3CRR sample), then perhaps the story of OUS would have been very different, or perhaps there 
would have been no such orientation-based unified scheme of powerful radio galaxies and quasars.
\begin{figure}
\scalebox{0.34}{\includegraphics{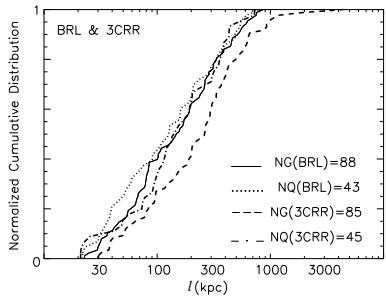}}
\caption{Normalized cumulative distribution of linear size ({\em l}) of RGs and quasars separately
for the 3CRR and BRL samples.}
\end{figure}

This brings us to a question whose scope goes much beyond OUS. Why do samples selected from two different parts of the sky but 
using almost the same selection criteria, show so different statistics of numbers as well as size distributions of 
radio galaxies and quasars? As already mentioned, the overall number ratios and redshift distributions do match in the two 
samples considered here. But as evident from Table 1, when we separate out the RGs and quasars in the samples, dividing 
them into different redshift bins, then the distributions in the two samples seem to differ qualitatively. 
While in the 3CRR sample OUS seems to be holding true in the intermediate 
redshift range $0.5\leq z<1$, in the BRL sample it seems to be so only at $z>1$. But then this is against the basic spirit of 
OUS where both RGs and quasars belong to the same parent population of radio sources, independent of redshift. Thus with its 
basic tenet, that the observed quasar sizes should be smaller than of RGs everywhere,   
having been violated, OUS seems to be precluded by the BRL sample.
 
There are other genuine differences which could not be due to some selection effects. For example, there are much larger 
number of CSSS in BRL (15\%) as compared to 5\% in the 3CRR sample, while the total numbers of sources in both samples are almost 
the same. Then there are 15\% FRIs in 3CRR catalogue, while in BRL they are only 7\% or so. Such differences need to be examined 
more thoroughly to find their statistical significance, as the radio source population should be similar in all sky regions 
(Copernican principle!), irrespective of whether or not the unified scheme holds good. 
Here we may point out that on a close examination, the 3CRR sample has shown large anomalies in the distribution of numbers 
and sizes of radio sources in different regions of the sky (Singal 2013). It remains 
to be seen how much contribution the OUS hypothesis has from these anomalies in the 3CRR sample. However, a 
similar anomaly is not seen in the BRL sample, therefore our present results are robust. 

\section{Conclusion}
We have shown that contrary to the expectations in OUS models, observed quasar sizes are not everywhere systematically smaller than 
those of galaxies in an independent but equivalent sample to that of the 3CRR. 
The absence of this foreshortening of the sizes of quasars as compared to those of RGs of similar flux densities 
or at similar redshifts, is inconsistent with the unified scheme models. 
While the two samples have roughly the same overall numbers of sources each, their distributions among RGs and quasars 
as a function of redshift do not seem to match, nor do the size distributions show similarities. 
It looks like that Barthel's observation that sizes and numbers of quasars were smaller than RGs 
was perhaps merely a statistical coincident or arose from some anomalies in the 3CRR sample, 
as similar things are not seen in other samples whether in a weaker MRC sample 
(Singal \& Singh 2013) or in a 3CRR-like strong-source BRL sample. Further there seems to be some large genuine differences 
in the two samples, selected using similar criteria, in two different regions of the sky that are rather intriguing and not in 
concordance with the conventional wisdom which expects an isotropic sky.


\begin{thebibliography}{}
\bibitem[]{} Antonucci, R., 1984, ApJ, 278, 499
\bibitem{7} Antonucci, R., 2012, Astr. Astrophys. Trans., 4, 557
\bibitem{2} Antonucci, R. R. J., \& Miller, J. S,. 1985, ApJ, 297, 621
\bibitem{1} Barthel, P. D., 1989, ApJ, 336, 606
\bibitem[]{} Best, P. N., R\"{o}ttgering, H. J. A., \& Lehnert, M. D., 1999, MNRAS, 310, 223
\bibitem[]{} Best, P. N., R\"{o}ttgering, H. J. A., \& Lehnert, M. D., 2000, MNRAS, 315, 21
\bibitem[]{} Best, P. N., Peacock, J. A., brookes, M. H., Dowsett, R. E., R\"{o}ttgering, H. J. A., Donlop, J. S., 
\& Lehnert, M. D., 2003, MNRAS, 346, 1021
\bibitem{14} Fanaroff, B. L., \& Riley, J. M., 1974, MNRAS, 167, 31P
\bibitem[]{} Hardcastle, M. J., Alexander, P., Pooley, G. G., \& Riley, J. M., 1998, MNRAS, 296, 445
\bibitem[]{} Hardcastle, M. J., Evans, D. A., \& Croston, J. H., 2009, MNRAS, 396, 1929
\bibitem[]{} Hine R. G., Longair M. S., 1979, MNRAS, 188, 111
\bibitem{9} Kapahi, V. K., Athreya, R. M., van Breugel, W., McCarthy, P. J., \& Subrahmanya, C. R., 1998a, ApJS, 118, 275
\bibitem{10} Kapahi, V. K., et al., 1998b, ApJS, 118, 327
\bibitem{3} Laing, R. A., Riley, J. M., \& Longair, M. S., 1983, MNRAS, 204, 151
\bibitem[]{} Laing R. A., Jenkins C. R., Wall J. V., Unger S. W., 1994, 
in Bicknell V., Dopita M. A., Quinn P. J., eds, ASP Conf. Ser. Vol. 54, The First Stromlo Symposium: 
The Physics of Active Galaxies. ASP, San Francisco, p. 201
\bibitem[]{} Large M. I., Mills B. Y., Little A. G., Crawford D. F., Sutton J. M., 1981, MNRAS, 194, 693
\bibitem[]{} Osterbrock D. E., Mathews W. G., 1986, ARA\&A, 24, 171
\bibitem[]{} Singal A. K., 1993, MNRAS, 262, L27
\bibitem{18} Singal, A. K., 2013, (submitted), arXiv:1305.4134
\bibitem[]{} Singal, A. K. \& Singh, R. L., 2013, ApJ, 766, 37 
\bibitem{19} Spergel, D. N, et al., 2003, ApJS, 148, 175
\end{thebibliography}
\end{document}